\documentclass[10pt,fleqn]{article}
\usepackage{amssymb}

\newcommand{\name}[1]{\begin{flushleft}
                       \LARGE \bf #1
                       \end{flushleft}\vspace{-3mm}}

\newcommand{\Author}[1]{\begin{flushleft}
                       \it #1 \end{flushleft}}

\newcommand{\Adress}[1]{\begin{flushleft}
                       \it #1 \end{flushleft}}

\newcommand{\be}{\begin{equation}}
\newcommand{\ee}{\end{equation}}
\newcommand{\ba}{\hspace*{-5pt}\begin{array}}
\newcommand{\ea}{\end{array}}
\newcommand{\p}{\partial}
\newcommand{\ds}{\displaystyle}

\newcommand{\pbf}[1]{\mbox{\mathversion{bold}$#1$}}

\begin{document}

\name{A relativistically invariant mass operator\par}

\medskip

\noindent{published in {\it Ukrainian Physics Journal}, 1968, {\bf
13}, N~3, P. 256--262.}

\Author{Wilhelm I. FUSHCHYCH\par}

\Adress{Institute of Mathematics of the National Academy of
Sciences of Ukraine, \\ 3 Tereshchenkivska  Street, 01601 Kyiv-4,
UKRAINE}

\noindent {\tt URL:
http://www.imath.kiev.ua/\~{}appmath/wif.html\\ E-mail:
symmetry@imath.kiev.ua}

\bigskip

\noindent In [1] it was shown how, for a given (discrete) mass
spectrum of elementary or hypothetical particles, it was possible
to construct a non-trivial algebra~$G$ containing a Poincar\'e
algebra $P$ as a subalgebra so that the mass operator, defined
throughout the space where one of the irreducible
representations~$G$
 is given, is self-conjugate and its spectrum coincides with the given mass spectrum.
Such an algebra was constructed in explicit form for the
nonrelativistic case, i.e., the generators were written for the
algebra. However, the problem of how to assign the algebra~$G$
constructively and determine an explicit form of the mass operator
in the relativistic case has remained unsolved.

In the present work we present a solution of this problem,
construct continuum analogs of the classical algebras~$U(N)$ and
$Sp(2N)$, and show that the problem of including the Poincar\'e
algebra can be formulated in the ``language'' of wave function
equations.

\renewcommand{\thefootnote}{\arabic{footnote}}
\setcounter{footnote}{0}

{\bf 1.} For simplicity, we will assume that there be given only
three particles with masses $m_1$, $m_2$ and $m_3$. $R_1$, $R_2$
and $R_3$ will represent the spaces in which irreducible
representations of the algebra $P$ are realized. The operator
$\left(P_\alpha^{(i)}\right)^2$ in these spaces is, as is well
known, a multiple of the unit operator:
\be
\left(P_\alpha^{(i)}\right)^2 R_i =m_i^2 R_i, \qquad i=1,2,3,
\quad \alpha=0,1,2,3. \ee We will designate by $R$ the linear sum
of these spaces\footnote{In the case of a real physical problem,
we should have taken the linear sum with certain  weights, the
squares of the moduli of which could be interpreted as the
probability  of finding the system in one or another of the
states.}. The operators of energy-momentum, angular momentum, and
square of the masses of the system, which may be in various
excited states in this space, take the form
\be
\ba{l} P_\alpha =P_\alpha^{(1)}F_{11} +P_\alpha^{(2)} F_{22}
+P_\alpha^{(3)} F_{33}, \vspace{1mm}\\ M_{\mu\nu}
=M_{\mu\nu}^{(1)} F_{11} +M_{\mu\nu}^{(2)}F_{22} +M_{\mu\nu}^{(3)}
F_{33}, \ea \ee
\be
M^2=\left(P_\alpha^{(1)}\right)^2 F_{11}+
\left(P_\alpha^{(2)}\right)^2 F_{22}+
\left(P_\alpha^{(3)}\right)^2 F_{33}, \ee where the $F_{ij}$
designate the squared three-rowed matrices in which unit operators
stand at the intersection of the $i$-th row and $j$-th column,
while all other elements are zero.

It is clear that in $R$ there are realized reducible
representations of the algebra $P$. However, relative to certain
sets of the operator $G$ this space may not have invariant
subspaces. Obviously, operators must appear in this set of the
type
\be
D=\left(\begin{array}{ccc} d_{11} & d_{12} & d_{13}\\ d_{21} &
d_{22} & d_{23}\\ d_{31} & d_{32} & d_{33} \ea \right), \ee where
at least one of the operators $d_{ij}$ $(i\not=j)$ is nonzero. In
order to solve our problem, these operators must be constructed in
explicit form.

For the determination of the explicit form of the operators use
will be made of the methods of the quantum theory of fields. The
vector $h_i\in R_i$ will be presented in the form~[2]
\be
h_i=\int d{\pbf k} \; F_i({\pbf k})a_i^+ ({\pbf k})|0\rangle. \ee
For simplicity of notation, we shall assume that the distribution
function $F_1({\pbf k})=F_2({\pbf k})=F_3({\pbf k})=F({\pbf k})$
and all particles are without spin. The generators of the
Poincar\'e algebra, expressed by the operators of creation and
annihilation, have the form~[3]
\be
\ba{l} \ds P_j^{(i)} =\int d{\pbf k}\; k_j a_i^+({\pbf k})
a_i({\pbf k}), \vspace{2mm}\\ \ds  P_0^{(i)}=\int d{\pbf k}\;
k_0^{(i)} a_i^+({\pbf k}) a_i({\pbf k}), \qquad
k_0^{(i)}=\sqrt{{\pbf k}^2+m_i^2}, \vspace{2mm}\\ \ds
M_{lr}^{(i)}=\frac i2 \int d{\pbf k} \left(k_lf_r^{(i)}({\pbf
k})-k_rf_l^{(i)}({\pbf k})\right), \qquad M_{0l}^{(i)}=\frac i2
\int d{\pbf k}\; k_0^{(i)}f_l^{(i)}({\pbf k}), \vspace{2mm}\\ \ds
f_l^{(i)}(\pbf{k}) =\frac{\p a_i^+({\pbf k})}{\p k_l} a_i({\pbf
k}) -a_i^+({\pbf k}) \frac{\p a_i({\pbf k})}{\p k_l}. \ea \ee

We can now write the explicit form of the operators
\be
d_{ij}=\int d{\pbf k}\; d{\pbf k}'\; F({\pbf k}) F'({\pbf k}')
\left\{ a_i^+({\pbf k}) a_j({\pbf k}')+ a_j^+ ({\pbf k}')
a_i({\pbf k})\right\}. \ee Obviously, the operators
$D_{ij}=d_{ij}F_{ij}$ will transform vectors from space $R_i$ to
$R_j$. Space $R$ is irreducible with respect to operators
$P_\alpha$, $M_{\mu\nu}$ and $D_{ij}$. This statement is a
consequence of the fact that the operators $D_{ij}$ transform a
given vector from $h\in R$, to vector $h_i\in R_i$, while, since
the subspace $R_i \subset R$ is noninvariant relative to these
operators then, by the same token, the irreducibility of the
representation $G$ in $R$ is shown.

The set of operators (5) and (6) (and their linear combinations)
form a Lie algebra in the case where they satisfy the Jacobi
identities. Calculating, for example, the commutators $[P_\alpha,
D_{ij}]_-$, $[P_\beta,[P_\alpha,D_{ij}]_-]_-$ etc., it is not
difficult to convince oneself that the operators derived from
these are not linear combinations of the operators $P_\alpha$,
$M_{\mu\nu}$ and $D_{ij}$, i.e., the set $G$ is an
infinite-dimensional Lie algebra. All elements of the algebra $G$
can be expressed explicitly by the operators $a_i^+({\pbf
k})a_j({\pbf k}')$, $a_j^+({\pbf k}')a_i({\pbf k})$, $[\p
a_i^+({\pbf k})/\p k_l]a_j({\pbf k}')$, $ a_l^+({\pbf k})[\p
a_r({\pbf k}')/ \p k'_i]$ and all possible products of these
operators. As will be shown below, these operators form a
continuous Lie algebra. In $R$ space the operators of hypercharge
and isospin have the form
\be
Y=j_1 F_{11}+j_2F_{22}+j_3F_{33}, \qquad J=i_1F_{11} +i_2 F_{22}+
i_3F_{33}, \ee where $j_1$, $j_2$, $j_3$ and $i_1$, $i_2$, $i_3$
are hypercharges and isospins of particles $m_1$, $m_2$  and
$m_3$. The formulas~(8) permit the expression of the
operator~$M^2$ by the operator of hypercharge and isospin. In our
case
\be
M^2 =a'E+b' Y +c'J, \ee where $E$  is the unit operator, and $a'$,
$b'$, $c'$  are arbitrary, generally speaking, constant
quantities. In the triplet representations of the algebra $G$,
which we considered, these quantities are uniquely determined by
the masses $m_1$, $m_2$ and $m_3$. In all other representations,
such uniqueness does not exist and hence formula~(9) will give a
mass relationship between the elementary particles. If the initial
particles have spin, then
\renewcommand{\theequation}{\arabic{equation}${}'$}
\setcounter{equation}{8}
\be
M^2 =aS+bY+cJ, \ee where $a$, $b$, $c$ are arbitrary numbers and
$S$ is the spin operator.

Other examples of infinite dimensional Lie algebras, containing
the algebra $P$, are considered in~[12].

\smallskip

\noindent {\bf Note 1.} It is well known that the masses of
elementary particles depend on spin, hypercharge, isospin, and
other quantum numbers; hence, for determining the mass operator,
one tends to express it by the operators of spin, hypercharge, and
isospin. It should be noted that, generally speaking, the mass
operator can always (in principle) be expressed by one operator.
In fact, let
\renewcommand{\theequation}{\arabic{equation}${}''$}
\setcounter{equation}{8}
\be
M^2 =f(A_1,A_2,\ldots, A_n), \ee where $A_1,A_2,\ldots,A_n$ are
mutually commuting self-adjoint operators, operating in a certain
separable Hilbert space. In agreement with Neiman's theorem~[4]
concerning creation operators, one can determine such a bounded
self-adjoint operator~$A$ in this space that
\[
A_n=\varphi_n(A).
\]
From this theorem it follows that
$M^2=f(A_1,A_2,\ldots,A_n)=\tilde f(A)$, i.e., the mass operator
can always be represented as a function of only one operator $A$
of a weakly closed ring. This attests to the fact that there
exists one universal quantum number, with the help of which it is
possible to explain the mass spectrum of elementary particles if
the explicit form of the function $\tilde f$ is known.

Since the mass operator in the approach arises from the same sort
of generator of the algebra~$G$ as, say, does the operator of
isospin or hypercharge, the formula ($9''$) may be viewed as an
equation of a hypersurface in a space of mutually commuting
operators. For such an interpretation of the mass formula ($9''$),
the generation opera\-tor~$A$  apparently plays the same role as
does time in classical mechanics (where the aggregate of all
trajectories lies on a certain manifold, in particular on a
surface $F(x,y,z)$ for which $x=x(t)$, $y=y(t)$, $z=z(t)$).

From the geometrical point of view the mass equations
\[
M=a+bS(S+1)
\]
for hadrons and
\[
M^2 =a^2 +b^2 S(S+1)
\]
for mesons represent ``trajectories'' (a parabola for hadrons and
hyperbola for mesons) of motion of the system, which can exist in
various mass and spin states.

The mass equations of Okubo,
\[
M=a+bY +c \{ J(J+1)-Y^2/4\}
\]
for hadrons and
\[
M^2 =a+bY +c \{J(J+1)-Y^2/4\}
\]
for mesons, represent a hyperbolic paraboloid and double poled
hyperboloid in an imaginary three-dimensional space $(M,Y,J)$.

In this manner, if we quantize the general equation for a
hyperbolic paraboloid:
\[
c/4 y^2 -cz^2 -cz -by +x-a=0,
\]
i.e., if in this equation we make the substitutions $x\to M$,
$y\to Y$, $z\to J$, then we will obtain the formula of Okubo for
hadrons. If with each multiplet we associate a definite
hypersurface, then various transitions of one multiplet to
particles of the same multiplet can be interpreted as ``motion''
or the given hypersurface. Transitions of particles of one
multiplet to particles of another multiplet may be considered as
transitions from one hypersurface to another. If to all
experimentally discovered hadrons (or bosons) is assigned a single
hypersurface, then all possible transitions of hadrons (bosons) to
hadrons (bosons) should be interpreted as ``motion'' on this
hypersurface, for which all quantum characteristics of the system
can change.

{\bf 2.} The characteristic special feature of problems concerning
the spectrum of atomic hydrogen and of a harmonic and anharmonic
oscillator, from the group theoretic standpoint, is that all these
problems can be solved by the method of embedding of the finite
dimensional Lie algebra, appropriate to groups of hidden symmetry,
in a broader but dimensionally finite Lie algebra~[5,~6]. However,
this statement does not depend on where the Hamiltonian is defined
--- in a Hilbert or in a vector space with indefinite metric.
Thus, for example, the problem of the spectrum of an
$N$-dimensional oscillator with complex ghosts can also be solved
by the method of embedding of a finite dimensional Lie algebra in
a finite dimensional Lie algebra\footnote{The question of
inclusion of an algebra of symmetry $U(2N)$ of such an oscillator
in a dynamic algebra will be considered in a subsequent paper.}.

From the above considerations (section 1) it follows that the
Poincar\'e algebra (relativistic case) can be included by a
nontrivial method only in the infinite di\-men\-sio\-nal Lie
algebra (the case of non-Lie algebras are not considered here).
This existing difference between the relativistic and
non-relativistic problem of the embedding of the Lie algebra can
be adequately explained in a natural manner. In quantum,
mechanics, as is well known, we always deal with finite numbers of
degrees of freedom. Transition to an infinite number of degrees of
freedom, apparently, implies a transition from a finite
dimensional Lie algebra to an infinite-dimensional one. We shall
expiate this statement with an example.

As was shown in [6], the space of states of an $N$-dimensional
harmonic oscillator realizes an irreducible representation of the
algebra $\overline U(N+1) \supset U(N)$. The generators of the
algebra $\overline U(N+1)$ satisfy the following commutation
relations:
\renewcommand{\theequation}{\arabic{equation}}
\setcounter{equation}{9}
\be
\left[ E_\rho^\lambda,
E_\varkappa^\sigma\right]_-=\delta_{\rho\sigma}E_\varkappa^\lambda-
\delta_{\varkappa\lambda}E_\rho^\sigma, \qquad \lambda, \rho,
\sigma, \varkappa=1,\ldots, N+1, \ee where
\be
\ba{l} \ds E_\mu^\nu =\frac 12 \left[a_\mu, a_\nu^+\right]_+,
\qquad \mu,\nu =1,\ldots,N, \vspace{2mm}\\ \ds E_{\mu}^{N+1} =g(H)
a_\mu^+, \qquad E_{N+1}^\mu =f(H) a_\mu, \qquad
E_{N+1}^{N+1}=h(H), \ea \ee
\be
H=\sum_{\mu=1}^N a_\mu^+ a_\mu, \qquad \left[ a_\mu,
a_\nu^+\right]_-=\delta_{\mu\nu}. \ee

If the $N$ number of the oscillators tends toward infinity, we
approach the infinite oscillator, but then the dynamic algebra of
an oscillator $\overline U(N+1)$ and the algebra of hidden
symmetry $U(N)$ go over into the infinite dimensional Lie algebra.
The algebra $Sp(2N)$ may be determined by an analogous method,
when $N\to \infty$.

For transition from quantum mechanics to the quantum theory of
fields, it is also necessary to let the volume in which the
oscillators are ``contained'' approach infinity~[2]. For such
passages to the limit, the operators $a_r$ and $a_s^+$ are
replaced by the general operators of annihilation $a({\pbf k})$
and creation $a^+({\pbf k})$, which satisfy the relations
\be
\left[ a({\pbf k}), a^+({\pbf k}')\right]_-=\delta({\pbf k}-{\pbf
k}'). \ee The dimensionally infinite algebra
${\mathop{U(N)}\limits_{N\to \infty}}$ for this case is naturally
associated with the continual algebra $U({\pbf k}, {\pbf k}')$,
the generators of which are the operators
\be
E({\pbf k}, {\pbf k}')= \frac 12 \left[ a({\pbf k}), a^+({\pbf
k}')\right]_+. \ee

It is not difficult to convince oneself that operators of the
form~(13) satisfy the following commutative relationships:
\be
\left[ E({\pbf k}, {\pbf k}'), E({\pbf q}, {\pbf q}')\right]_-=
\delta({\pbf k}-{\pbf q}') E({\pbf q}, {\pbf k}')- \delta({\pbf
k}'-{\pbf q}) E({\pbf k}, {\pbf q}'). \ee Further, let us
construct the algebras $U_N({\pbf k}, {\pbf k}')$ and
$Sp_{2N}({\pbf k}, {\pbf k}')$. Consider the set of operators:
\be
E_\mu^\nu({\pbf k}, {\pbf k}')= \frac 12 \left[ a_\mu({\pbf k}),
a_\nu^+({\pbf k}')\right]_+, \qquad \mu,\nu=1,\ldots, N, \ee
\be
E_{\mu\nu}({\pbf k}, {\pbf k}')=a_\mu({\pbf k})a_\nu ({\pbf k}'),
\qquad E^{\mu\nu}({\pbf k}, {\pbf k}')=a_\mu^+({\pbf k})a^+_\nu
({\pbf k}'), \ee where
\be
\left[ a_\mu({\pbf k}), a_\nu^+({\pbf
k}')\right]_-=\delta_{\mu\nu} \delta({\pbf k}-{\pbf k}'). \ee
Taking into account (18), it can be shown that
\be
\left[ E_\mu^\nu({\pbf k}, {\pbf k}'), E_\alpha^\beta({\pbf q},
{\pbf q}')\right]_-= \delta_{\mu\beta} \delta({\pbf k}-{\pbf q}')
E_\alpha^\nu({\pbf q}, {\pbf k}')- \delta_{\nu\alpha} \delta({\pbf
q}-{\pbf k}') E_\mu^\beta({\pbf k}, {\pbf q}'), \ee
\be
\left[ E_{\mu\nu}({\pbf k}, {\pbf k}'), E_{\alpha\beta}({\pbf q},
{\pbf q}')\right]_-=0, \ee
\be
\left[ E_\mu^\nu({\pbf k}, {\pbf k}'), E_{\alpha\beta}({\pbf q},
{\pbf q}')\right]_-= -\delta_{\nu\beta} \delta({\pbf k}'-{\pbf
q}') E_{\alpha\mu}({\pbf q}, {\pbf k})- \delta_{\nu\alpha}
\delta({\pbf q}-{\pbf k}') E_{\beta\mu}({\pbf q}', {\pbf k}), \ee
\be
\left[ E_\mu^\nu({\pbf k}, {\pbf k}'), E^{\alpha\beta}({\pbf q},
{\pbf q}')\right]_-= \delta_{\mu\beta} \delta({\pbf k}-{\pbf q}')
E^{\alpha\nu}({\pbf q}, {\pbf k}')+ \delta_{\alpha\mu}
\delta({\pbf k}-{\pbf q}) E^{\beta\nu}({\pbf q}', {\pbf k}'), \ee
\be
\ba{l} \left[ E_{\mu\nu}({\pbf k}, {\pbf k}'),
E^{\alpha\beta}({\pbf q}, {\pbf q}')\right]_-= \delta_{\nu\alpha}
\delta({\pbf k}'-{\pbf q}) E_\mu^\beta({\pbf k}, {\pbf q}')+
\delta_{\mu\alpha} \delta({\pbf k}-{\pbf q}) E^\beta_\nu({\pbf
k}', {\pbf q}')+ \vspace{2mm}\\ \qquad  +\delta_{\nu\beta}
\delta({\pbf k}'-{\pbf q}') E^\alpha_\mu({\pbf k}, {\pbf q})+
\delta_{\mu\beta} \delta({\pbf k}-{\pbf q}) E^\alpha_\nu({\pbf
k}', {\pbf q}), \ea\hspace{-10.3pt} \ee
\be
\left[ E^{\mu\nu}({\pbf k}, {\pbf k}'), E^{\alpha\beta}({\pbf q},
{\pbf q}')\right]_-=0. \ee

The set of operators $\left\{ E_\mu^\nu({\pbf k}, {\pbf
k}')\right\}$, satisfying the relations~(19) form a continuous Lie
algebra $U_N({\pbf k}, {\pbf k}')$. The set of operators $\left\{
E_\mu^\nu({\pbf k}, {\pbf k}'), E^{\mu\nu}({\pbf q}, {\pbf
q}')\right\}$, satisfying the relationships (19)--(24), form the
continuous Lie algebra $Sp_{2N}({\pbf k}, {\pbf k}')$.

Utilizing the commutating relations (19)--(24) it is possible to
show that the elements from $Sp_{2N}({\pbf k}, {\pbf k}')\supset
U_N ({\pbf k}, {\pbf k}')$ satisfy the Jacobi identity. Since
elements of the algebra $U_N({\pbf k}, {\pbf k}')$ depend
continuously on the variables ${\pbf k}$ and ${\pbf k}'$, it is
then possible to formally determine the derivative
\be
\frac{\p E_\mu^\nu({\pbf k}, {\pbf k}')}{\p k_i} \equiv
A^i_{\mu\nu}({\pbf k}, {\pbf k}'), \qquad \frac{\p E_\mu^\nu({\pbf
k}, {\pbf k}')}{\p k'_j} \equiv B^j_{\mu\nu}({\pbf k}, {\pbf k}'),
\qquad i,j=1,2,3. \ee

Taking into account (19), it is not difficult to establish that
\be
\ba{l} \ds \left[ A^i_{\mu\nu}({\pbf p}, {\pbf p}'),
A^j_{\alpha\beta}({\pbf q}, {\pbf q}')\right]_-= \vspace{2mm}\\
\ds \qquad =\delta_{\mu\beta}\frac{\p \delta ({\pbf p}-{\pbf
q}')}{\p p_i} A^i_{\alpha\nu}({\pbf q}, {\pbf p}')-
\delta_{\alpha\nu}\frac{\p \delta ({\pbf q}-{\pbf p}')}{\p q_j}
A^i_{\mu\beta}({\pbf p}, {\pbf q}'), \ea \ee
\be
\ba{l} \ds \left[ B^i_{\mu\nu}({\pbf p}, {\pbf p}'),
B^j_{\alpha\beta}({\pbf q}, {\pbf q}')\right]_-= \vspace{2mm}\\
\ds \qquad =\delta_{\mu\beta}\frac{\p \delta ({\pbf p}-{\pbf
q}')}{\p q'_j} B^i_{\alpha\nu}({\pbf q}, {\pbf p}')-
\delta_{\alpha\nu}\frac{\p \delta ({\pbf q}-{\pbf p}')}{\p p'_i}
B^j_{\mu\beta}({\pbf p}, {\pbf q}'), \ea \ee
\be
\ba{l} \ds \left[ E_\mu^\nu({\pbf p}, {\pbf p}'),
A^i_{\alpha\beta}({\pbf q}, {\pbf q}')\right]_-= \vspace{2mm}\\
\ds \qquad =\delta_{\mu\beta} \delta ({\pbf p}-{\pbf q}')
A^i_{\alpha\nu}({\pbf q}, {\pbf p}')- \delta_{\alpha\nu}\frac{\p
\delta ({\pbf q}-{\pbf p}')}{\p q_i} E^\beta_\mu({\pbf p}, {\pbf
q}'), \ea \ee
\be
\ba{l} \ds \left[ E_\mu^\nu({\pbf p}, {\pbf p}'),
B^j_{\alpha\beta}({\pbf q}, {\pbf q}')\right]_-= \vspace{2mm}\\
\ds \qquad =\delta_{\mu\beta}\frac{\p \delta ({\pbf p}-{\pbf
q}')}{\p q'_j} E_\alpha^\nu({\pbf q}, {\pbf p}')-
\delta_{\alpha\nu}  \delta ({\pbf q}-{\pbf p}')
B^j_{\mu\beta}({\pbf p}, {\pbf q}'), \ea \ee
\be
\ba{l} \ds \left[ A^i_{\mu\nu}({\pbf p}, {\pbf p}'),
B^j_{\alpha\beta}({\pbf q}, {\pbf q}')\right]_-= \vspace{2mm}\\
\ds \qquad =\delta_{\mu\beta}\frac{\p^2 \delta ({\pbf p}-{\pbf
q}')}{\p p_i \p q'_j} E_\alpha^\nu({\pbf q}, {\pbf p}')-
\delta_{\alpha\nu} \delta ({\pbf q}-{\pbf p}') \frac{\p^2
E_\mu^\beta({\pbf p}, {\pbf q}')}{\p p_i \p q'_j}. \ea \ee

Analogously, the relation may be established also for the
derivatives
\[
\frac{\p E_{\mu\nu}({\pbf k}, {\pbf k}')}{\p k_i}, \qquad \frac{\p
E_{\mu\nu}({\pbf k}, {\pbf k}')}{\p k'_j}, \qquad \frac{\p
E^{\mu\nu}({\pbf k}, {\pbf k}')}{\p k'_i}, \qquad \frac{\p
E^{\mu\nu}({\pbf k}, {\pbf k}')}{\p k_j}.
\]

From the relations (26), (28) and (27), (29) it can be seen that
the set of operators $\left\{ E_\mu^\nu({\pbf p}, {\pbf p}'),
A^i_{\mu\nu}({\pbf q}, {\pbf q}')\right\}$ and $\left\{
E_\mu^\nu({\pbf p}, {\pbf p}'), B^i_{\mu\nu}({\pbf q}, {\pbf
q}')\right\}$ also form a continuous Lie al\-geb\-ra.

For consideration of the continuous Lie algebras, we may
introduce, by analogy to the classical Lie algebra theory, the
concepts of the universal enveloping Lie algebra, the center,
Casimir operators, etc. It is clear that all these concepts
require refinement from the mathematical point of view since, so
far as we know, such Lie algebras are not considered in the
mathematical literature. As regards the problem of classification
and formulation of all irreducible representations of the algebra
$Sp_{2N}({\pbf k}, {\pbf k}')$, it leads, as can be seen from
relations~(16) and~(17), to the problem of the description of all
unitary non-equivalent commutation relations~(18). This last
problem, as is known, has not been solved up to the present time.

With the operators $E_\mu^\nu ({\pbf p}, {\pbf p}')$, apparently,
one cannot directly associate certain physical quantities (energy,
momentum, angular momentum, etc.). However, the in\-teg\-ral
operators derived from these operators, i.e., operators of the
type
\[
E_\mu^\nu =\int d{\pbf p}\; d{\pbf p}'\; f_\mu^\nu ({\pbf p},
{\pbf p}') E_\mu^\nu ({\pbf p}, {\pbf p}'),
\]
as can be seen from Sec. 1, can be assigned definite physical
meanings.

It is possible to display other continuous Lie algebras. Thus, for
example, the operators
\[
\left\{ E_\mu^\nu ({\pbf k}, {\pbf k}') , E_{\mu_1\mu_2\ldots
\mu_n} ({\pbf p}_1, \ldots, {\pbf p}_n)= a_{\mu_1}({\pbf
p}_1)a_{\mu_2}({\pbf p}_2)\ldots a_{\mu_n}({\pbf p}_n)\right\}
\]
or
\[
\left\{ E_\mu^\nu ({\pbf k}, {\pbf k}') , E^{\mu_1\mu_2\ldots
\mu_n} ({\pbf p}_1, \ldots, {\pbf p}_n)= a^+_{\mu_1}({\pbf
p}_1)a^+_{\mu_2}({\pbf p}_2)\ldots a^+_{\mu_n}({\pbf p}_n)\right\}
\]
also form continuous Lie algebras.

{\bf 3.} In [7] it was shown that the set of infinitesimal
operators of homogeneous Lorentz group $O(3,1)$ and operators
$L_\mu$, entering into the relativistic equation
\be
\left( L_\mu\frac{\p}{\p x_\mu} +\varkappa\right) \Phi (x_0, {\pbf
x})=0, \qquad \mu=0,1,2,3 \ee form a Lie algebra, which is an
isomorphous set of infinitesimal operators of the de Sitter group
$O(4,1)$. The function $\Psi(x_0,{\pbf x})$ for a Lorentz
transformation is trans\-for\-med according to the representation
$R=\sum\limits_{i=1}^n \oplus R_i^{l_0^i,l_1^i}$, where
$\left(l_0^i,l_1^i\right)$ are pairs of numbers to which are given
the irreducible representations of $O(3,1)$. Since the generators
of group $O(3,1)$ and operators  $L_\mu$ transform one solution of
Eq.(31) to another solution, it is clear that in all solution sets
of~(31) there are realized irreducible representations of group
$O(4,1)$. Since $\Phi(x_0,{\pbf x})$ pertains to a apace which is
a linear sum of spaces in which is realized the irreducible
representation $O(3,1)$, then, obviously, the spectrum of the
Casimir operators,
\[
K_1 =-\frac 12 M_{\mu\nu} M_{\mu\nu}, \qquad K_2 =-\frac 14
\varepsilon_{\mu\nu\rho\sigma} M_{\mu\nu} M_{\rho\sigma}, \qquad
\mu,\nu, \rho,\sigma=0,1,2,3
\]
in this space will be discrete.

On the basis of the above it is natural to propose the following
problem: to formulate an equation for the wave function $\Psi$
which would be invariant relative
 to the Poincar\'e group and in all sets of solutions (solution space)
of this equation of the spectrum of Casimir operators,
\be
P^2=P_\mu P_\mu, \qquad W^2 =W_\alpha W_\alpha, \qquad
W_\alpha=\frac 12 \varepsilon_{\alpha\beta\gamma\delta} P_\beta
M_{\gamma\delta} \ee would be discrete.

For the solution of this problem we will use one of the results of
Foldy~[8]. In~[8] it was shown that with each irreducible unitary
representation of the Poincar\'e group with mass m and spin s
there can be associated a Schr\"odinger equation
\be
H\Psi(x_0, {\pbf x})=i\frac{\p \Psi (x_0, {\pbf x})}{\p t}, \ee
where $H=({\pbf P}^2 +m^2)^{1/2}$ and $\Psi(x_0,{\pbf x})$ is the
$(2s+1)$-component wave function, quadratically integrable over
the space variables. The question of the uniqueness of such
correspondence (i.e., the question of possible existence of
another equation which would also express the free motion of a
relativistic particle with mass $m$ and spin $s$) is left open
in~[8].

The single ambiguity, which apparently arises from the
establishment of this correspondence, is tied to the extraction of
the square root of the operator ${\pbf P}^2+m^2$. Actually there
is no such ambiguity, since the operator ${\pbf P}^2+m^2$ is
positive, and by virtue of theorems~[10] the square root of a
positive self-adjoint operator is uniquely determined. This is
proof in itself that the stated correspondence is isomorphic.

If the Hamiltonian in Eq.(34) is expressed in the form
\[
\widetilde H=\sqrt{{\pbf P}^2 +M^2},
\]
where ${\pbf P}^2 =P_1^2+P_2^2+P_3^2$, and $M^2$ is the operator
determined by formula~(3) it can then be seen that~(34) is a
natural generalization of the relativistic Eq.(33) (in which the
constant value $m^2$ is replaced by the operator $M^2$) in the
case where the particle can take on various mass states.

In this manner, every relativistic equation expressing a free
particle of mass $m$ and spin $s$ is unitarily equivalent to
Eq.(33) $(H>0)$.

Since the Casimir operators $P^2$ and $W^2$ enter the theory on
equal terms, then we may use the operator $W^2$ to obtain the
equation of motion of a free particle. In this case, the equation
which, generally speaking, is unitarily equivalent to Eq.(33), has
the form
\renewcommand{\theequation}{\arabic{equation}${}'$}
\setcounter{equation}{32}

\be
\sqrt{W^2+m^2 s(s+1)} X({\pbf x},t) =W_0X({\pbf x},t), \ee i.e.,
between $X$ and $\Psi$,  there exists the coupling $X=V\Psi$,
where $V$ is the isometric operator.

Establishment of isomorphism between the Schr\"odinger equations
and the ir\-re\-du\-cib\-le unitary representations of the
Poincar\'e group permits the writing of the equation which would
have the above state properties. This equation has the form
\renewcommand{\theequation}{\arabic{equation}}
\setcounter{equation}{33}
\be
\widetilde H \widetilde \Psi_+ =i\frac{\p \widetilde \Psi_+}{\p
t}, \ee where
\[
\widetilde H=\!\!\left( \begin{array}{ccc} \!\!\sqrt{{\pbf
P}^2+m_1^2}\!\! & 0 &0 \vspace{1mm}\\ 0 & \!\!\sqrt{{\pbf
P}^2+m_2^2}\!\! & 0 \vspace{1mm}\\ 0 & 0& \!\!\sqrt{{\pbf
P}^2+m_3^2}\!\! \ea \right)\!\!, \qquad \widetilde \Psi_+
=\!\!\left( \!\!\begin{array}{c} \Psi_+^{m_1,s_1}(x_0,{\pbf x})
\vspace{1mm}\\ \Psi_+^{m_2,s_2}(x_0,{\pbf x}) \vspace{1mm}\\
\Psi_+^{m_3,s_3}(x_0,{\pbf x}) \!\! \ea \right)\!\! .
\]
The plus sign means that the sign value of the supplementary
Casimir operator (the sign of the energy) for the Poincar\'e
group~[9] for these solutions is equal to~$+1$.

The Schr\"odinger equation which would also be invariant under
time reflection has the form
\be
H' X=i\frac{\p X}{\p t}, \ee where
$
H'=\left( \begin{array}{cc} \widetilde H & 0 \\ 0 & \widetilde H
\ea \right)$,
$X=\left( \begin{array}{c} \widetilde \Psi_+\\ \widetilde \Psi_-
\ea \right).
$

In conclusion, let us note that, in agreement with the theorems of
O'Raifeltaigh~[13] in the space of the solutions of Eq.(34) one
cannot realize an irreducible representation of a finite
dimensional Lie algebra which would contain the Poincar\'e algebra
as a subalgebra.

If  Eqs. (31) and (33) are considered equivalent (the unitary
equivalence is con\-st\-ruc\-ted only for equations describing
particles with spin $1/2$), then the formula for $\Phi'(x_0,{\pbf
x})$, expressing motion of a particle which may be in various mass
states, has the same formal appearance as the equations for
elementary particles. However, the quantity $\varkappa$ is then
not a constant but a variable, taking on the following values:
\be
\varkappa=\pm m_1 \lambda_1, \pm m_2\lambda_2,  \pm m_3 \lambda_3,
\ldots, \ee where $m_i^2=p_0^2-{\pbf p}^2$ and $\lambda_i$  is
some real nonzero eigenvalue of the operator $L_0$. In~[15] it is
shown that only for such values of $\varkappa$ do Eqs.(31) have
nonzero plane wave solutions.

The relation (36) can be written in the form of the mass formula:
\renewcommand{\theequation}{\arabic{equation}${}'$}
\setcounter{equation}{35}
\be
M=\varkappa L_0^{-1}. \ee

The operators which transform solutions of Eq.(31) with fixed mass
to solutions which have a different mass are constructed from
creation and annihilation operators by an analogous method (as in
Sec.~1).

\smallskip

\noindent {\bf Note 2.} Equation (31), as was shown in [11],
excluding the Dirac equation, cannot be reduced by the unitary
representations of the Foldy--Woythysen type to a Schr\"odinger
equation. Consequently, the function $\Phi(x_0,{\pbf x})$,
strictly speaking, is not a wave function of a particle with fixed
mass $m$ and spin $s$.

The construction of a non-trivial theory of interaction based on
Eq.(35), i.e., the introduction of potential in~(35), by excluding
those theories which with the help of unitary representations
reduce to free particles (or as is generally stated, to the theory
of free quasiparticles)~[3], meets with difficulties in
practice~[14].

From the previous considerations, with every elementary particle
there is as\-so\-ci\-a\-ted a space $R_i$, in which is realized an
irreducible representation of algebra $P$. A~particle which can be
found in various excited states is associated with space $R$
 which is a linear sum of the spaces $R_i$.
The inadequacy of such an approach lies in the fact that all
elementary particles are considered as stable, and consequently
possessing definite mass. Actually, a definite mass to these
resonances cannot be ascribed, since particles are then nonstable.

To account for this fact, it is sufficient in the above mentioned
considerations to change the linear sum to the linear integral:
\renewcommand{\theequation}{\arabic{equation}}
\setcounter{equation}{36}
\be
R=\int \oplus R(m) g(m), \ee where the metric $g(m)$ is
concentrated on the set composed of one or more points (depending
on how many stable particles) and nonoverlapping intervals $[m'_i,
m''_i]$.

A more expanded formulation of equation~(37) has the appearance
\be
R=R^{s_0}(m_0) \oplus \sum_{i=1}^m R_i, \ee
\be
R_i=\int \oplus R^{s_i}(m) f^{s_i}(m) dm, \ee where $R^{s_i}(m)$
is the space in which is realized the irreducible representation
of algebra~$P$ with mass $m$ and spin $s_i$; the function
$f^{s_i}(m)$, nonzero only in the interval $(m'_i,m''_i)$,
characterizes the ``smearing'' (indeterminacy) of the mass of a
resonance. If in~(39) we replace $f^{s_i}(m)$ by a delta function,
then $R$, as before, will be a linear sum of spaces $R_i$.

The operator $\left(P_\alpha^{(i)}\right)^2$ in $R_i$ is
determined in the following manner:
\renewcommand{\theequation}{\arabic{equation}${}'$}
\setcounter{equation}{0}
\be
\left(P_\alpha^{(i)}\right)^2R_i =\int \oplus
\left(P_\alpha^{(i)}\right)^2 R^{s_i} (m) f^{s_i}(m) dm=\int
\oplus m^2 R^{s_i} (m) f^{s_i}(m) dm. \ee The operators
$P_\alpha^{(i)}$, $M_{\mu\nu}^{(i)}$, $M^2$, $P^2$ can be
determined by an analogous method. A~more detailed presentation of
results obtained by taking account of ``smearing'' of the
resonances will be given in another paper.

\medskip

\begin{enumerate}

\footnotesize

\item Fushchych W.I., {\it Ukr. Fiz. Zh.}, 1967, {\bf 12}, 741.

\item Bogolyubov N.N., Shirkov D.V., Introduction to the Theory of Quantized Fields, Wiley, 1959.

\item Fushchych W.I., {\it Ukr. Fiz. Zh.}, 1967, {\bf 12}, 1331.

\item Neiman I., Mathematical Bases of Quantum Mechanics,  Nauka, 1964 (in Russian).

\item Barut A., {\it Phys. Rev.}, 1965, {\bf 139}, 1433;\\
Malkin I.A., Man'ko V.I., 1965, {\bf 2}, N~5, 230, {\it Sov. Phys.
-- JETP. Lett.}, 1965, {\bf 2}, 146.

\item Hwa R., Nuyts J., {\it Phys. Rev.}, 1966, {\bf 145}, 1188.

\item Fushchych W.I., {\it Ukr. Fiz. Zh.}, 1966, {\bf 11}, 907.

\item Foldy L., {\it Phys., Rev.}, 1956, {\bf 102}, 568.

\item Shirokov Yu.M., {\it Zh. Eksp. Teor. Fiz.}, 1957, {\bf 33}, 1196.

\item Riss F., SekefaI'vi-Nad' B., Lectures on Functional Analysis, IL, 1954 (in Russian).

\item Jordan T., Mukunda N., {\it Phys. Rev.}, 1963, {\bf 132}, 1842.

\item Formanek J., {\it Czech. J. Phys. B}, 1966, {\bf 16}, 1;\\
Votruba I., Gavlichek M., Physics of High Energies and the Theory
of Elementary Particles, Kiev, Naukova Dumka, 1967, P.~330.

\vspace{-2mm}

\item O'Raifertaigh L., {\it Phys. Rev. Letters}, 1965, {\bf 14}, 575.

\item Schweber S., Introduction to the Relativistic Quantum Theory of Fields, Harper, 1961.

\item Gel'fand I.M., Michlos R.A., Shapiro E.Ya.,
Representations of Rotation and Lorentz Groups, Moscow, 1958.

\item Mettews P.T., Salam A., {\it Phys. Rev.}, 1958, {\bf 112}, 283.
\end{enumerate}

\end{document}